%% file: primary_tex_file.tex
\documentclass[runningheads]{llncs}
\usepackage[T1]{fontenc}
\let\oldproof\proof
\let\oldendproof\endproof

\usepackage{graphicx}
\usepackage{booktabs}
\usepackage{array}
\usepackage{geometry}
\usepackage{framed}
\usepackage{amsmath}
\usepackage{xcolor}

\let\proof\oldproof
\let\endproof\oldendproof

\begin{document}
\title{Diffusion Mechanism Design in Tree-Structured Social Network}
 \author{Feiyang Yu}
 
\institute{Duke University}
%
\maketitle              
\begin{abstract}
We design a fixed-price auction mechanism for a seller to sell multiple items in a tree-structured market. The buyers have independently drawn valuation from a uniform distribution, and the seller would like to incentivize buyers to invite more people to the auction. We prove that our mechanism is individual rational, and incentivize compatible with regard to the buyers' action. Furthermore, we show the approximation ratio of our mechanism to the optimal fixed-price auction in two ways, theoretically and via Monte-Carlo simulation, and show a high practical ratio. Finally, we discuss several factors affecting the behavior of our mechanism and its feasibility in reality.

\end{abstract}

\input{introduction}

\input{model}

\input{mechanism}
\input{evaluation}
\input{related_work}
\input{discussion}

\section*{Acknowledgement}
We thank Dengji Zhao and Tianyi Zhang (author of \cite{zhang2021fixed}) for formulating the research idea of diffusion mechanism design in social networks.

\bibliographystyle{splncs04}
\bibliography{ref}
\end{document}

%% file: introduction.tex
\section{Introduction}

People have been running auctions for the need to buy and sell items since ancient times. Fixed-price auctions are the most widely used selling method due to their simplicity and ease of use~\cite{mathews2004impact}. In a traditional fixed-price auction, the size of an auction (i.e., the number of participants) greatly affects the seller's revenue~\cite{krishna2009auction}, but also limited by the seller's social connection. To be more specific, the seller wants to expand the set of buyers for a higher revenue, but it is traditionally difficult and expensive to reach more people beyond who they already know. Nowadays, with rapidly advancing technology, social platforms such as Facebook and X (formerly Twitter) has greatly changed how people interact with each other. Therefore, mechanism design on social network has drawn attention from researchers in various fields~\cite{domingos2001mining}. 

One notable change brought by modern social media is the significantly low cost on spreading information. An individual can easily find new buyers in an "indirect" way through a friend on social network. An intuitive approach inspired by such fact is to incentive existing buyers for diffusing sales information to other people they know. Many prior researchers~\cite{li2017mechanism,li2020information,li2022diffusion,li2024diffusion,li2024double,zhang2021fixed,zhang2020redistribution} have taken this approach and proposed partial solutions. Li et al.~\cite{li2017mechanism} propose a mechanism applicable when the entire network structure and buyers' valuations are visible to the seller; while Zhang et al.~\cite{zhang2021fixed} introduce a fixed-price diffusion mechanism for a single item.

However, one of the challenges in fixed-price mechanisms is the difficulty in approximating optimal revenue. Several prior searches have studied different optimal conditions and bounds of fixed-price mechanisms compared to Myerson's optimal auction mechanism~\cite{alaei2019optimal,babaioff2017posting}. Therefore, it is also crucial to numerically evaluate the designed fixed-price mechanism's revenue.

The goal of this paper is to explore how to help sellers improve their revenue in a multi-item auction scenario by incentivizing more buyers to join. Our mechanism is more general than~\cite{zhang2021fixed} since it applies to the sales of multiple items, and more privacy-preserving than~\cite{li2017mechanism,li2022diffusion,li2024diffusion} since our mechanism does not require knowledge of the buyers' valuations.

\textbf{Contributions.} We summarize our contributions as follows:
\begin{enumerate}
    \item We formally formulate the problem of selling items in social networks into a mechanism design problem.
    \item We design a fixed-price diffusion mechanism that is feasible, individual rational, and incentive compatible.
    \item We prove the strict lower bound and calculate the average approximation ratio of our mechanism.
\end{enumerate}

\textbf{Outline.} Section~\ref{sec:model} formally define several important terms and the model of the problem. Section~\ref{sec:mechanism_design} presents the design of our fixed-price mechanism. Section~\ref{sec:evaluation} evaluates our mechanism by its approximation ratio. Section~\ref{sec:related_work} provides an overview of related literature. We discuss a few known limitations and potential future work in Section~\ref{sec:discussion}, and concluse in Section~\ref{sec:conclusion}.

%% file: model.tex
\section{The Model}
\label{sec:model}

We consider the market where a seller $S$ would like to sell $m$ identical items via a social network. We abstract the social network as a graph $G = (V, E)$. The set of nodes $V$ consists of the root node $S$ representing the seller, and $n$ other nodes representing potential buyers in the market. Edges in $E$ represent the "information flow": for each edge $(i, j) \in E$, $i$ is able to diffuse the sales information to $j$. In our model, we assume the social network graph $G$ has the structure of a tree.

For simplicity, we denote the set of buyers $K = V - \{S\}$. Each buyer $i \in K$ would like to purchase at most 1 item, and has a \textit{private} valuations $v_i \geq 0\ (i \in \{1,2,...,n\}$. We assume all buyers' valuations are independently drawn from the uniform distribution over the interval $[0, 1]$~\cite{myerson1981optimal}.

We then formalize our model and give some necessary definitions. Let buyer $i$'s type be $\theta_i$, while $(\theta_i)_{i \in K}$ is the set of all buyers' types. In this model, a buyer's decision is either i) informing their neighbors (successors) $(\theta_{-i})$, or ii) taking no action ($\emptyset$). The \textbf{type profile} $a(\theta_i)_{i \in K}$ is the set of all buyers' profiles. For the aforementioned decisions, we denote \textbf{action} $a'_i := \theta'_i || \emptyset$. $a'_i = \theta'_i$ means buyer $i$ spreads the sales information, while $a'_i = \emptyset$ means the opposite. Finally, we denote the \textbf{action profile} $a' = (a_i)_{i \in K}$ and $a'_{-i} = (a_j)_{j \in (K - \{a_i\})}$.

We borrow many definitions from~\cite{zhang2021fixed} because of the similar problem settings.

\begin{definition}[Feasible action]
Given buyers' type set ${\theta_1, \theta_2, ..., \theta_k}$, $\forall i \in K$: $a_i' \neq \emptyset$ and $\theta_i' \subseteq \theta_i$, action profile $a'$ is \textbf{feasible} if and only if $i$ obeys action set $a_{-i}$
\end{definition}

A feasible action profile $a'$ guarantees that a node (buyer) only participates in the auction if another node informs them. 

\begin{definition}[Mechanism]
A mechanism $(\pi, p)$ consists of a set of allocation strategies $\pi = (\pi_i)_{i \in K}$ and a set of payment functions $p = (p_i)_{i \in K}$. Denote $\mathcal{F}(a)$ as the set of all feasible action profiles given type profile $a$. 
\begin{itemize}
    \item $\pi_i : \mathcal{F}((n_i)_{i \in K}) \to \{0, 1\}$ is called buyer $i$'s \textbf{allocation strategy}.
    \item $p_i: \mathcal{F}((n_i)_{i \in K}) \to [0, 1]$ is called buyer $i$'s \textbf{payment function}.
\end{itemize}
\end{definition}

\begin{definition}[Feasible allocation]
	If $\forall a' \in F_n$,
	\begin{itemize}
		\item $\forall i \in K$, if $a_i'=\emptyset$, then $\pi_i(a') = 0$.
		\item $\sum_{i \in K}\pi_i(a') \in \{0, 1, ..., m\}$.
	\end{itemize}
	then allocation strategy $\pi$ is \textbf{feasible}.
\end{definition}

Simply, a feasible allocation will not allocate items to buyers not participating in the auction, and allocate an item to at most one buyer.

\begin{definition}[Utility]
\label{definition:utility}
Given action $a'$ and mechanism $(\pi, p)$, buyer $i$'s utility is defined as
	\[
	u_i(a', (\pi,p)) = \pi(a')v_i - p_i(a')
	\]
\end{definition}

\begin{definition}[Individual Rationality]
If $\forall i \in K, \forall a' \in  F((n_i)_{i \in K}), u_i(a', (\pi, p)) \geq 0$, then mechanism $(\pi, p)$ is \textbf{individual rational}.
    
\end{definition}

\begin{definition}[Diffusion Incentive Compatibility]
If $\forall i \in K, $ for $a_i$ that $a_i' = \theta_i$, $\forall a'' \in F((n_i)_{i \in K})$, $u_i(a', (\pi, p)) \geq u_i(a'', (\pi, p))$ always holds, then mechanism $(\pi, p)$ is \textbf{diffusion incentive compatible}.
\end{definition}

Notice that the definition for \textit{diffusion incentive compatibility} (DIC) differs from the standard definition of \textbf{incentive compatibility}~\cite{roughgarden2010algorithmic}. DIC indicates the action of diffusing information is the dominant strategy of all buyers.

In the next section, we present our mechanism and prove that it is i) individual rational, and ii) diffusion incentive compatible.

%% file: mechanism.tex
\section{Mechanism Design}
\label{sec:mechanism_design}

Before formally introducing our mechanism, we need to define a few more symbols. Given the tree-structured social network graph $G$, let $d_i$ be the depth of buyer $i$ in the graph, and $PH_i$ be the set of all nodes (buyers) on the path from root (seller) $S$ to $i$. Denote the set of the $x$ immediate successors of root $s$ as $n_s = \{1, ..., x\}$.

The sub-tree rooted at node $i \in n_s$ is called a \textbf{branch}. Denote the set of all nodes in the branch rooted at $i$ as $B_i$ and $k_i = |B_i|$. Further denote $k_{-i} = k - k_i$. Finally, the \textbf{reward factor} $\alpha$ is a tunable parameter to adjust the reward for spreading the sales information. Our mechanism is designed as the following:

\begin{framed}
\textbf{Multi-item Diffusion Mechanism}\\
\rule{\textwidth}{0.5pt}\\
1. The seller $S$ calculates $p_i(a')=\frac{1}{1+\frac{k_{-i}}{x}}^{\frac{x}{k_{-i}}}$ for each branch $B_i$.\\
2. $S$ broadcast price $p_i(a')$ to all buyers in $B_i$, calculate the number of allocated items $a_i = \lfloor \frac{m\cdot k_i}{k} \rfloor$, and set $m = m - a_i$.\\
3. Any buyers who have a higher valuation than the broadcast price $p_i(a')$ express their interest in purchasing.

\begin{itemize}
    \item \textbf{Allocation:} In each branch $B_i$, $S$ picks $a_i$ buyers who express interest, with the smallest depth and the most successors (with random tie-breaking). If there are not enough interested buyers, $S$ allocates items to all buyers.
\end{itemize}
\begin{itemize}
    \item \textbf{Payment:} If an item is allocated to buyer $j \in B_w$, then for all buyers $l \in B_w$, their payments are:
	\[ p_l(a') =\begin{aligned}  \begin{cases}
	p_w(a') \quad & \pi_l(a') = 1\\
	- p_w(a')\alpha (\frac{1}{2})^{d_l} \quad & \pi_l(a') = 0 \& l \in PH_j \\
	0 \quad & otherwise
	\end{cases} \end{aligned}
	\]
\end{itemize}
\end{framed}

The design of this mechanism include one intuitive but yet challenging consideration. To determine a proper payment function for a \textit{branch} that both improves the seller's revenue and keeps the mechanism truthful. The optimal price in a single-item fixed-price mechanism without diffusion given number of buyers $N$ can be calculated to be \cite{zhang2021fixed}:
\[p_0 = (\frac{1}{1+n})^{\frac{1}{n}}\] 

Since it's mathematically difficult to derive an analytical expression of the optimal price for the multiple-item case, our mechanism approximates the single-item optimal price $p_0$. However, $n$ in the formula must be substituted by a carefully chosen value that i) approximates $n$ well, and ii) is independent of $n$ (to keep the mechanism truthful). In practice, we choose $\frac{k_{-i}}{x}$ (average branch size excluding $B_i$) as the heuristic to obtain the payment function $p_i(a')$, and show such a heuristic works well in Section~\ref{sec:evaluation}.

Being a generalized version of the single-item mechanism in prior work \cite{zhang2021fixed}, our mechanism also enjoys several desirable properties.

\begin{theorem}[Individual Rationality] Our mechanism is individual rational.
\end{theorem}
\textit{Proof.} We consider the utility of $i$ in two stages: the first stage where buyers consider their utility bidding for the item, and the second state where buyers receive their rewards for diffusing the sales information. We denote the branch $i$ belongs to as $X$. In the first stage, we consider the utility resulting from the transaction of the item:
\begin{itemize}
    \item If $\pi_i(a') = 1$: since $i$ is allocated an item, $i$'s valuation $v_i$ must be higher than the announced price to them, and therefore have a positive utility.
    \item If $\pi_i(a') = 0$: $i$ is not allocated an item, so their utility is 0 (Definition~\ref{definition:utility}).
\end{itemize}

In the second stage, we consider the utility resulting from the diffusion action. For each successful item allocation to $j \in X$:
\begin{itemize}
    \item If $i \in PH_j$: $i$ receives $p_w(a')\alpha (\frac{1}{2})^{d_l}$ (which is positive).
    \item If $i \notin PH_j$: $i$'s utility is 0.
\end{itemize}

Since either $\pi_i(a') = 1$ or $\pi_i(a') = 0$, and either $i \in PH_j$ or $i \notin PH_j$, $i$'s utility is the sum of that in the two stages before. The final utility is non-negative, given the utility in each stage is non-negative.

\begin{theorem}[Diffusion Incentive Compatible] Our mechanism is diffusion incentive compatible.
\end{theorem}
\textit{Proof.} Similarly, we consider the utility of a buyer $i$ in the same two stages. In the first stage,
\begin{itemize}
    \item If $\pi_i(a') = 1$: $i$'s utility is $v_i - p_w(a')$. Since $p_w(a')$ is independent of $k_w$, whether $i$ chooses to diffuse the sales information does not impact their utility from the transaction. On the other hand, not diffusing the information lowers the number of successors of $i$ in the graph. It therefore lowers the chance they are allocated an item (according to our allocation rule), and their utility decreases by doing so. So, diffusing the information is $i$'s dominant strategy.
    \item If $\pi_i(a') = 0$: $i$'s utility is 0. 
\end{itemize}

In the second stage, notice that the rewards to $i$ from diffusion monotonically increase with regard to the size of the branch. $i$ also increases their chances of getting the rewards by diffusion (effectively adding nodes to the sub-tree) and therefore increases their overall utility by doing so. Combining the two stages together, $i$'s utility always improves when they choose to diffuse the information. Therefore, diffusing is their dominant strategy, and our mechanism is \textit{diffusion incentive compatible}.

\begin{theorem}[Time Complexity]
    The running time complexity of our mechanism is $O(|V|+|E|)$.
\end{theorem}
\textit{Proof.} The seller needs to know the structure of the graph $G$, specifically the number of nodes in each branch, the depth and the number of successors of each node. This can be obtained by running a single Breadth-First-Search on $G$, which has the time complexity of $O(|V|+|E|)$.

%% file: evaluation.tex
\section{Evaluation}
\label{sec:evaluation}

In this section, we evaluate the efficacy of our mechanism in two ways, by i) proving the theoretical lower bound, and ii) evaluating the average approximation ratio by Monte Carlo experiments. 

\subsection{Baselines}
\label{subsec:baselines}
In order to evaluate our mechanism, we first define two simple auction mechanisms, the non-diffusing fixed-price auction (\textbf{baseline}) and the optimal fixed-price auction (\textbf{optimal}). Consider the following multiple-item fixed-price auction:

\begin{framed}
\textbf{Fixed-price Auction}\\
\rule{\textwidth}{0.5pt}
1. The seller sets a fixed price $p \in [0, 1]$, and informs all buyers.\\
2. Any buyers whose valuation of 1 item are greater than $p$ expresses their interest in purchasing.

\begin{itemize}
    \item \textbf{Allocation:}  The seller randomly picks $m$ buyers who express interest if there are at least $m$ such buyers. Otherwise, the seller chooses all interested buyers.
\end{itemize}
\begin{itemize}
    \item \textbf{Payment:} If there is any buyer allocated an item, their payment is $p$.
\end{itemize}
\end{framed}

In our model, we define the baseline auction as the fixed-price auction among the seller's direct neighbors and its revenue as $R_0$. Similarly, we define the optimal auction to be the fixed-price auction among all potential buyers in the graph and its revenue as $R_{opt}$. Intuitively, $R_0$ represents the revenue of \textbf{not} applying our mechanism, and $R_{opt}$ represents the maximal possible revenue to run a fixed-price auction among all participants.

\subsection{Lower bound}

\begin{theorem}
Our mechanism's approximation ratio has a lower bound of 0.25.
\end{theorem}
Imagine the intuitive worst-case scenario to apply our mechanism where the seller $S$ would like to sell only 1 item, and there are $n \rightarrow \infty$ buyers directly connected to $S$. Our diffusion mechanism would randomly pick one buyer at a fixed price, effectively conduct a fixed-price auction to sell 1 item to 1 buyer. The revenue of such an auction given price $p$ can be calculated as:
\[
R_D = p(1-p)
\]

$R_D$ is maximized when $p$ is set to 0.5, which results in a revenue of 0.25. On the other hand, since the optimal auction is to sell 1 item among $n$ buyers, the chance of selling it at $p=1$ approaches 1 as $n \rightarrow \infty$, which brings the optimal revenue to 1. Therefore, in the worst-case scenario, our mechanism achieves the approximation ratio of $0.25/1=0.25$.

\subsection{Average approximation ratio}
In addition to lower bound in theory, we also evaluate our mechanism by its practical average approximation ratio. Notice an important observation on the reward factor $\alpha$: when $\alpha \rightarrow 0$, the "sub-auction" in a branch is effectively a fixed-price auction among all nodes in that branch. Hence, given number of items $m$, number of potential buyers $n = |V|-1$,  branch sizes $K = \{k_1, k_2, ..., k_n\}$, item allocations to corresponding branches $A = \{a_1, a_2, ..., a_n\}$ and finally the fixed price $p$, the revenue of our mechanism can be calculated as
\[
 R_D = \sum_{k_i \in K, a_i \in A}\sum^{k_i}_{n=1} {a_i \choose n} (1-p)^n p^{a_i-n} + k_i p {a_i \choose n}(1-p)^n p^{a_i-n}
\]

There are two major challenges to calculate the average ratio: i) it is mathematically hard to obtain an analytical solution for $p$ with regard to $K$ and $A$, and ii) it is computationally expensive to iterate all possible tree structures even when we fix $m$ and $n$ (Cayley's formula~\cite{cayley1881analytical} shows there are $n^{n-2}$ possible tree structures). 

Alternatively, we run a set of Monte Carlo experiments to calculate an approximated average ratio in the following steps:
\begin{enumerate}
    \item[(1)] Generate a random integer sequence of a chosen length $|V|$.
    \item[(2)] Compute a corresponding tree structure using the Prüfer algorithm~\cite{prufer1918neuer}. 
    \item[(3)] Calculate the revenue of the non-diffusing fixed-price auction $R_0$.
    \item[(4)] Calculate the revenue of the optimal fixed-price auction $R_{opt}$.
    \item[(5)] Calculate the revenue of our mechanism $R_D$.
\end{enumerate}

Step (2) ensures the generated tree obeys a uniform distribution. Our results for different tree sizes $n$ are summarized in Table~\ref{table:NTable}. The results show that the approximation ratio of our algorithm ($R_D/R_{opt}$) grows and approaches 1 as $n$ grows. In comparison to the baseline auction, our mechanism results in a consistent increase in revenue as well.

We further evaluate the impact of $n$ and $k_i$. We generate several sets of trees with expected size of 10,000, each with different means of $n$ and $k_i$. The results are summarized in Table~\ref{table:NKTable}. It shows that our mechanism better works on trees with more branches.

\begin{table*}[ht!]
\centering
\begin{tabular}{|>{\centering\arraybackslash}p{2cm}|>{\centering\arraybackslash}p{3cm}|>{\centering\arraybackslash}p{3cm}|} 
\hline
\textbf{n} & \textbf{$R_D / R_0$} & \textbf{$R_D / R_{opt}$} \\ 
\hline
10      & 1.41 & 0.43 \\ 
\hline
20      & 1.70 & 0.52 \\ 
\hline
40      & 1.98 & 0.60 \\ 
\hline
100     & 2.35 & 0.71 \\ 
\hline
500     & 2.48 & 0.84 \\ 
\hline
1000    & 2.32 & 0.88 \\ 
\hline
10000   & 2.08 & 0.91 \\ 
\hline
100000  & 2.28 & 0.97 \\ 
\hline
\end{tabular}
\caption{Comparison of $R_D / R_0$ and $R_D / R_{opt}$ for different values of N}
\label{table:NTable}
\end{table*}

\begin{table*}[ht]
\centering
\begin{tabular}{|>{\centering\arraybackslash}p{2cm}|>{\centering\arraybackslash}p{3cm}|>{\centering\arraybackslash}p{3cm}|>{\centering\arraybackslash}p{3cm}|} 
\hline
\textbf{Mean of $n$} & \textbf{Mean of $k_i$} & \textbf{$R_D / R_0$} & \textbf{$R_D / R_{opt}$} \\ 
\hline
5   &   200 &   2.63    & 0.89  \\
\hline
10  &   100 &   2.36    & 0.87  \\
\hline
20  &   50  &   2.09    & 0.83  \\
\hline
50  &   20  &   1.73    & 0.77  \\
\hline
100 &   10  &   1.45    & 0.65  \\
\hline
\end{tabular}
\caption{Comparison of $R_D / R_0$ and $R_D / R_{opt}$ for different values of $n$ and $k_i$}
\label{table:NKTable}
\end{table*}

%% file: related_work.tex
\section{Related Work}
\label{sec:related_work}
There have been prior work on social network mechanism designs, and designing diffusion mechanisms.

\textbf{Social network mechanism designs.} Mechanism designs on social networks have been more and more popular in recent years. Some related work include~\cite{li2017mechanism,hargreaves2019fairness,kawasaki2020strategy}. While the prior works also study on social network, our work focuses on a specific problem, multiple-item auctions in a specific graph structure with a more extensive evaluation.

\textbf{Diffusion mechanism designs.} There have been several prior works on the topic of diffusion mechanism design. Notable works include~\cite{li2022diffusion,li2024double,li2024diffusion,zhang2021fixed}. \cite{li2022diffusion} proposes the information diffusion mechanism (IDM), and generalizes it to the mechanism family critical diffusion mechanisms (CDM) to solve the single-item diffusion auction problem. \cite{li2024diffusion} further generalizes the work by adding consideration of transaction cost. However, their work requires prior knowledge of the buyers' valuations, and is relatively more expensive to compute. In contrast, our mechanism is more privacy-preserving and efficient as it does not require buyers' valuations and has a lower time complexity.

Our work is a generalized extension of the single-item diffusion mechanism design proposed in \cite{zhang2021fixed}. Compared to the prior work, our study has a more general application scenario, includes an extensive evaluation of the approximation ratio, and has a significantly less computational complexity (compared to running the single-item mechanism multiple times).

%% file: discussion.tex
\section{Discussion}
\label{sec:discussion}

We discuss several known limitations of our study in this section.

\textbf{Limitation of graph structures.} In this paper, we limit the scope of our mechanism to only tree-structured graphs. Past study has shown that the structure of social networks can be very flexible~\cite{jackson2006economics,jackson2008social,jackson2011overview}. Even in the simplified case of spreading information, it is natural for the "information flow" graph to form cycles, or even nested cycles. Prior work has shown the transformation from a directed-acyclic-graph (DAG) to a tree~\cite{zhang2021fixed} without impacting the mechanism's desirable properties, but the real-world social networks are much more complicated. An intuitive challenge would be to decide the reward in a cycle and in the case of diamond dependencies while keeping the diffusion incentive compatibility property of the mechanism. It remains a open question to design a mechanism that fits in more general social network structures.

\textbf{Impact of $\alpha$.} The reward factor $\alpha$ is an vital part of our mechanism. While we assume $\alpha$ to be a tiny non-negative value to ensure the properties of our mechanism, $\alpha$ taking value of any positive real number would have an impact on the overall mechanism and its outcome. One important case to consider is the reward for diffusing may exceed the utility of the transaction when the valuation is close enough to the announced fixed price. The revenue decrease for introducing $\alpha$ is also interesting to study. An estimation can be calculated based on the fact that a random node in a uniformly random tree of size $n$ has the expected depth $E[D] \approx \sqrt{\pi n}$. If we assume all buyers have an equal chance to win an item (which is a very rough estimation due to the diffusion part of our mechanism), the revenue of the buyers could decrease by a portion of $\sum_{i=1}^{\lfloor \sqrt{\pi n} \rfloor} \alpha^i$. The value would be roughly 1\% when $\alpha =0.01$ and $n=100$. We consider the impact of $\alpha$ to be a minor yet interesting consideration, and part of the future work.

\textbf{Improvement of the lower bound.} While we show that our mechanism only has a 0.25 approximation ratio in the very worst case, such worst case is very easy to avoid since the seller knows the size of branches before running the auction. The seller can choose a heuristic to adjust price and item allocation. For example, if the seller already knows there are far more branches than items to sell, they can "merge" branches to improve their revenue. As we show our mechanism has a satisfying average approximation ratio, it is still an open question how to efficiently improve the bad case revenues.

\section{Conclusion}
\label{sec:conclusion}
In this paper, we present a multiple-item mechanism design in tree-structured graphs that helps sellers improve their revenue by incentivizing buyers to "diffuse" the information. The designed mechanism has several desirable properties, including individual rationality and diffusion incentive compatibility. In addition, we show the approximation ratio of our mechanism has a lower bound of 0.25, and a practical average of 0.97 when the social network is large enough. In the future, we plan to further generalize our mechanism.